\documentclass{kapproc} 

\usepackage{procps} 

\usepackage[dvips]{graphicx}

\upperandlowercase

\kluwerbib 

\begin{document}

\articletitle{Evidence for large superhumps in TX Col and V4742 Sgr}


\author{Alon Retter}
\affil{School of Physics\\
University of Sydney, 2006\\
Australia}
\email{retter@Physics.usyd.edu.au}

\author{Alexander Liu}
\affil{Norcape Observatory\\
PO Box 300\\
Exmouth, 6707\\
Australia}
\email{asliu@onaustralia.com.au}

\author{Marc Bos}
\affil{Mt Molehill Observatory\\
83a Hutton Street\\
Otahuhu, Auckland \\
New Zealand}
\email{cushla@kiwilink.co.nz}

\begin{abstract}
Since the discovery of the largest positive superhump period in TV Col
(6.4 h), we have started a program to search for superhumps in cataclysmic 
variables (CVs) with large orbital periods. In this work, we summarize 
preliminary results of our observations of TX Col and V4742 Sgr. TX Col 
is an intermediate polar with a 5.7-h orbital period. V4742 Sgr is a 
recent (2002) nova with no known periods. CCD unfiltered continuous 
photometry of these two objects was carried out during 56 nights (350 
hours) in 2002-2003. The time series analysis reveals the presence of 
several periods in both power spectra. In TX Col, in addition to the 
orbital period of 5.7 h, we found peaks at 7.1 h and 5.0 h. These 
are interpreted as positive and negative superhumps correspondingly, 
although the effects of the quasi-periodic oscillations at $\sim$2 h 
(which may cause spurious signals) were not taken into consideration.
In the light curve of V4742 Sgr two long periods are detected -- 6.1 and 
5.4 h as well as a short-term period at 1.6 h. This result suggests that 
V4742 Sgr is an intermediate polar candidate and a permanent superhump 
system with a large orbital period (5.4 h) and a superhump period excess 
of 13\%. If these results are confirmed, TX Col and V4742 Sgr join TV Col
to form a group of intermediate polars with extremely large superhump
periods. There seems to be now growing evidence that superhumps can occur 
in intermediate polars with long orbital periods, which is very likely 
inconsistent with the theoretical prediction that superhumps can only 
occur in systems with mass ratios below 0.33. Alternatively, if the mass 
ratio in these systems is nevertheless below the theoretical limit, they 
should harbour undermassive secondaries and very massive white dwarfs, 
near the Chandrasekhar limit, which would make them excellent candidates 
for progenitors of supernovae type Ia. 
\end{abstract}


\section*{Introduction}

Binary systems often show quasi-periodicities a few percent longer than
their orbital periods. They are understood as the beat periods between the
orbital period and the apsidal precession of the accretion disc. For
historical reasons they are known as positive superhumps. The positive
superhumps obey a nice relation between the superhump period excess over
the orbital period and the orbital period (Stolz \& Schoembs 1984;
Patterson 1999). Negative superhumps, quasi-periodicities a few percent
shorter than the orbital periods, are explained by the beat periods
between the orbital period and the nodal precession of the accretion disc.
The negative superhumps follow a somewhat similar relation between the
superhump period deficit over the orbital period and the orbital period
(Patterson 1999, see also Retter 2002). Superhumps are important as the 
binary mass ratio can be estimated from the observed difference between 
the superhump and orbital period.

According to theory (Whitehurst \& King 1991; Murray 2000) precessing 
accretion discs can occur only in binaries with small mass ratios 
(q=$M_{2}/M_{1}\leq$1/3). In CVs, systems with longer orbital periods 
have larger separations and their secondaries are thus more massive
since they have to fill their Roche Lobes. There is a small scatter on
the mass of the primary white dwarf and therefore, the limit on the 
mass ratio is translated into orbital periods shorter than about 3-4~h. 
TV~Col, with an orbital period of 5.5~h and a negative superhump of 
5.2~h has been an unusual case. Retter et al. (2003) found another 
period, 6.4~h, in existing data of this object and confirmed it by 
further observations. It is naturally understood as a positive 
superhump. These results raised the question whether TV~Col is unique. 
This work shows that it is almost certainly not.

\section{Observations, Analysis and Results}
\subsection{TX~Col}

Photometric unfiltered CCD observations of TX~Col, a 5.7-h orbital 
period CV (Buckley \& Tuohy 1989; Norton et al. 1997), were carried out 
using a 0.3-m Meade LX200 telescope and a CCD in Norcape Observatory, 
Exmouth, Australia during 22 nights from December 2002 until February 
2003 and in 7 nights in December 2003. No filter was used. Fig. 1 shows 
the light curve of this system.


\begin{figure}[]
\vskip.2in
\centerline{\includegraphics[width=3in]{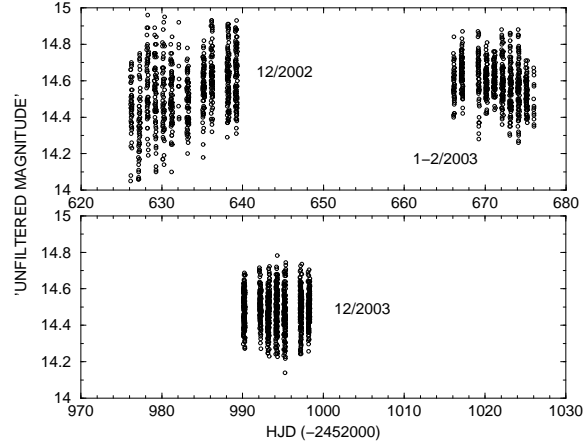}}
\caption{The light curve of TX Col during 22 nights in 2002-2003 and 7
nights in 2003.}
\end{figure}

Fig. 2 presents the power spectra (Scargle 1982) of these observations. 
Note that for the first observing season (December 2003 -- February 2002) 
runs shorter than 5 h were rejected and the mean was subtracted from each 
night. In the power spectrum of the second season (December 2003) no 
de-trending method was use. The peaks at the right hand-side of the diagram 
(which are more evident in the observations of the second season) correspond 
to the spin period (1911 s, 45.2 d$^{-1}$) and its beat with the orbital 
period (2106 s, 41.0 d$^{-1}$). The light curve also has quasi-periodic 
oscillations at $\sim$2 h ($\sim$12 d$^{-1}$).

\begin{figure}[]
\vskip.2in
\centerline{\includegraphics[width=3in]{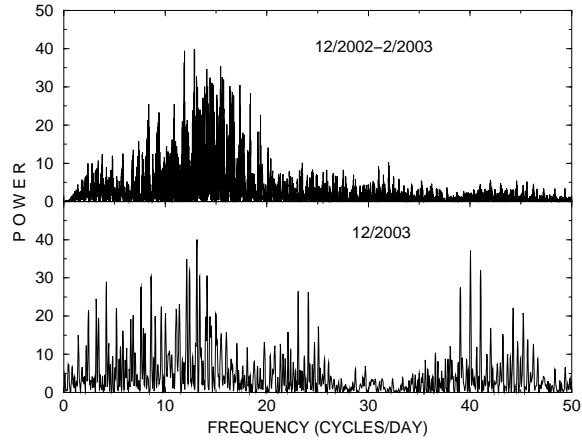}}
\caption{Power spectra of two observing seasons of TX Col in 2002-2003. 
Top panel: December 2002 -- February 2003. Bottom panel: December 2003. 
See text for more details.}
\end{figure}

In Fig. 3 we show the frequency interval 1-7 d$^{-1}$. The data from the
first season (top panel) show two groups of peaks in addition to the
orbital period (f$_{3}$). The peak-to-peak amplitudes of these signals 
are about 0.1 mag. We propose that the 5.2-h period (f$_{1}$) is a 
negative superhump and that the 7.0-h period (f$_{2}$) -- a positive 
superhump. `a$_{i}$' (i=1-3) represent 1-d$^{-1}$ aliases of `f$_{i}$' 
correspondingly. The data from the second observing season confirms the
presence of the 7.0-h period (f$_{2}$).

\begin{figure}[]
\vskip.2in
\centerline{\includegraphics[width=3in]{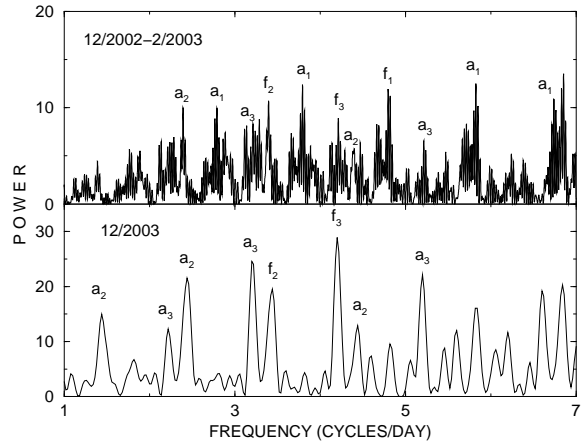}}
\caption{Same as Fig. 2 for the frequency interval 1-7 d$^{-1}$.
f$_{3}$ is the known orbital period (5.7 h). In the power spectrum of 
the first observing run there are two additional groups of peaks 
centered around f$_{1}$ (5.2 h) and f$_{2}$ (7.0 h) which we interpret 
as negative and positive superhumps respectively. f$_{2}$ also 
appears in the data of the second season. See text for more details.}
\end{figure}

\subsection{V4742 Sgr}

Fig. 4 displays the photometric unfiltered CCD observations of Nova 
V4742 Sgr 2002/2 taken during 26 nights in May-July 2003. The nova 
decayed by about 0.6 mag during the observing run. The nightly light 
curves show erratic behaviour, which is typical of data which are 
modulated with several periods and which is similar to the light 
curves of TV~Col and TX~Col. Two long runs are shown as examples in 
Fig. 5. 

\begin{figure}[]
\vskip.2in
\centerline{\includegraphics[width=3in]{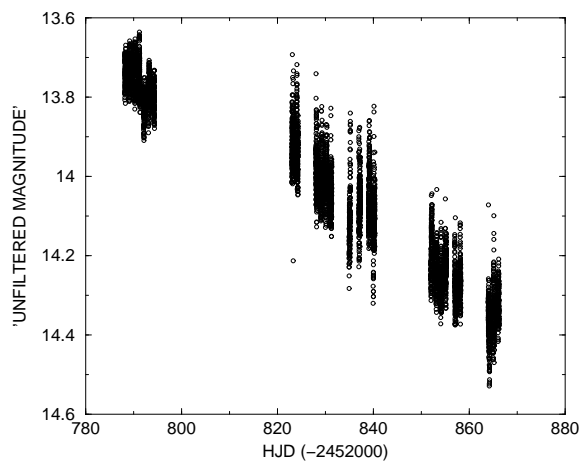}}
\caption{The unfiltered light curve of Nova V4742 Sgr 2002/2 during
26 nights in May-July 2003.}
\end{figure}

\begin{figure}[]
\vskip.2in
\centerline{\includegraphics[width=3in]{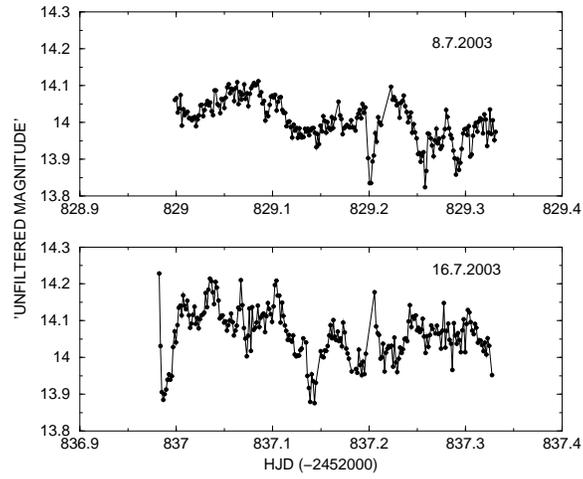}}
\caption{Two sample light curves of V4742 Sgr. Top panel: July 8th, 2003.
Bottom panel: July 16th, 2003}
\end{figure}

The power spectrum of the whole data (after subtracting the long-term 
trend and the mean from each night) showed several peaks around 6 h, 
however it was somewhat noisy due to the presence of short runs. 
Therefore, runs shorter than 0.29 d ($\sim$7 h) were rejected. Figs. 6 
and 7 present the power spectrum of the remaining 15 nights. Similar
to the data of TX Col, the power spectrum of V4742 Sgr displays a 
complicated multi-periodic structure. We could identify two long-term 
frequencies in the data -- 3.96 and 4.48 d$^{-1}$ , which correspond 
to 6.1 and 5.4 h respectively. The peak-to-peak amplitudes of these signals
are about 0.05 mag. The difference between the two periods ($\sim$13\%) 
would fit a positive superhump excess if we interpret the 5.4~h peak as 
the orbital period and the 6.1 h peak as a positive superhump. 

The presence of the 5.4~h period was confirmed by CCD unfiltered 
observations using a 0.25-m telescope in Mt Molehill Observatory, 
Auckland, New Zealand during 5 nights (25 h) in July-August 2003. 


\begin{figure}[]
\vskip.2in
\centerline{\includegraphics[width=3in]{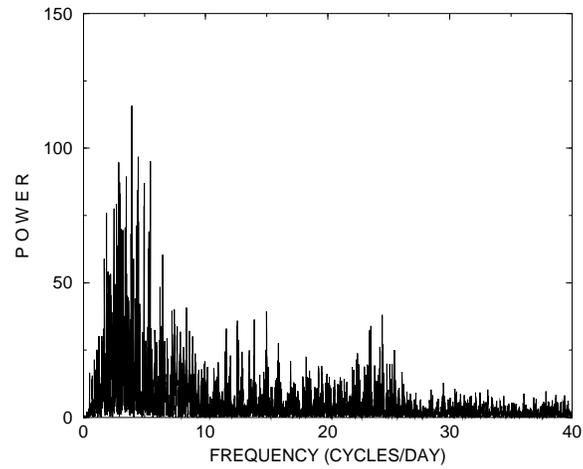}}
\caption{Power spectrum of the 15 longest runs of Nova V4742 Sgr 2002/2.
It shows several peaks. See also Fig.~7.}
\end{figure}

The power spectrum of V4742 Sgr also shows several peaks at shorter 
periods. This structure suggests that the nova is an intermediate polar 
system with a spin period of 1.6 h or 59 min. Confirmation by X-ray
observations is naturally required for this suggestion.

\begin{figure}[]
\vskip.2in
\centerline{\includegraphics[width=3in]{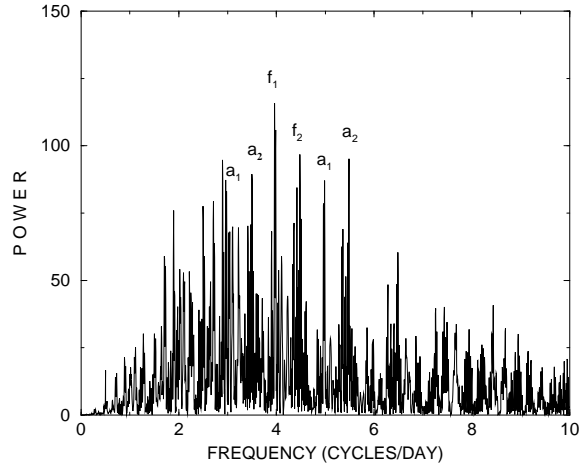}}
\caption{A zoom of Fig.~6 into the frequency interval 0-10 d$^{-1}$.
The power spectrum shows two long-term periods and their 1-d$^{-1}$
aliases. We suggest that the 5.4-h peak (f$_{2}$) is the orbital period 
and that the 6.1-h peak (f$_{1}$) is a positive superhump. a$_{i}$ 
(i=1,2) correspond to 1-d$^{-1}$ aliases of f$_{i}$ respectively. In 
addition, the peaks at longer frequencies (Fig. 6) support an 
intermediate polar model for this system.}
\end{figure}

\subsection{Tests}

The presence of the periodicities in the data of TX Col and V4742 Sgr 
was checked by several tests. We calculated the power spectrum of the 
airmasses; we subtracted one period and 
checked whether the other/s disappear from the power spectrum of the 
residuals; we planted the period/s in the data and checked its/their 
aliases. Simulations were also carried out to check the significance 
of one period in the presence of the other/s. The data were divided into 
subsections and the power spectra of the runs were compared. We also 
checked different de-trending methods (and in particular subtracting the 
trend from each night, which is especially relevant for the decaying 
nova, V4742 Sgr).

These tests support the above findings. We note, however, that for 
TX Col we did not try to estimate the influence of the quasi-periodic 
oscillations. Quantifying and simulating this effect is extremely hard. 
The presence of quasi-periodic oscillations may cause spurious signals. 
In the case of V4742 Sgr, this effect (if exists at all) is weak and 
cannot form the observed strong signals unless the periods themselves 
represent quasi-periodic oscillations.

Another warning comment is that the detected signals are near the length
of the nightly runs. We tried to overcome this problem by rejecting short 
runs, however, for TX Col observations shorter than 5 h (shorter than
the suggested periods) were still included. For V4742 Sgr, nights shorter 
than 7 h were rejected, therefore the remaining nights are longer than 
the proposed periods. This means that our results of V4742 Sgr stand on 
a safer ground. Anyway, it is recommended to confirm these results by 
further, multi-site observations.

\section{Discussion}

The analysis of the data of TX Col and V4742 Sgr reveals strong evidence 
that they have several periods in their light curves. The uncertainties 
in these findings were outlined in the previous section. Assuming that 
these results are real, and adopting our interpretation of the periods 
of TX Col and V4742 Sgr, we plotted in Fig. 8 the extension of the 
relation for positive superhumps to long periods. TV Col and V4742 Sgr 
obey the relation while TX Col deviates from it having a superhump 
period excess somewhat larger than the predicted value. This diagram 
suggests that superhumps may be common in CVs with orbital periods up 
to about 6 h (or even larger).

\begin{figure}[]
\vskip.2in
\centerline{\includegraphics[width=3in]{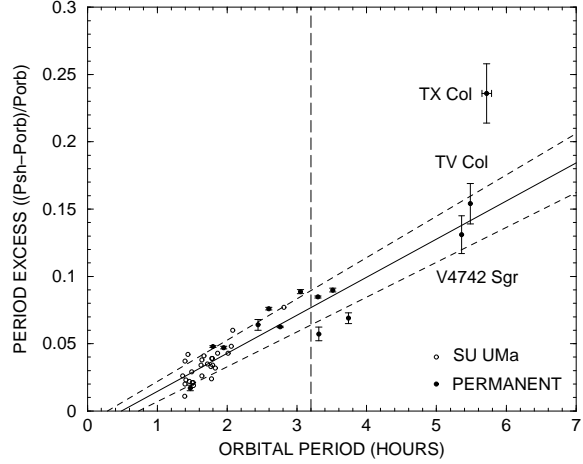}}
\caption{The relation between the positive superhump excess and the 
orbital period. TV Col and V4742 Sgr obey the relation while TX Col 
somewhat deviates from it. Note that the values for V4742 Sgr and 
TX Col still require confirmation.}
\end{figure}

\section{Summary}

This work gives evidence that TX Col and V4742 Sgr have several large 
periods and thus supports the idea that superhumps can be found in CVs 
with large orbital periods. We note, however, that the light curve of 
TX Col shows quasi-periodic oscillations with large amplitudes that 
complicate the analysis and may cause spurious signals. The data 
of V4742 Sgr do not show this behaviour, however, to firmly state that
the nova is a permanent superhump system with a large orbital period
its orbital period should be confirmed by a radial velocity study. 
The analysis of the photometric observations of V4742 Sgr also 
indicate that it may be an intermediate polar system. Thus, we feel 
that there is growing evidence that superhumps can occur in CVs with 
orbital periods of 5-6 h. This almost certainly means that they have 
mass ratios larger than the theoretical limit of 0.33. The reason for 
this behaviour may be that they all all intermediate polars. The 
presence of superhumps with large periods may be alternatively 
understood if the systems have undermassive secondary stars and 
massive primary white dwarf. In this case, these systems are excellent 
candidates for supernovae type IA.

We encourage observers to look for superhumps is CVs with large
orbital periods including dwarf novae above the period gap.

\begin{acknowledgments}

AR is supported by a grant from the Australian Research Council.

\end{acknowledgments}

\end{document}